\documentclass{article}

\usepackage{arxiv}
\usepackage{graphicx}

\usepackage[utf8]{inputenc} 
\usepackage[T1]{fontenc}
\usepackage{url}
\usepackage{ifthen}
\usepackage{cite}
\usepackage[cmex10]{amsmath} 
\allowdisplaybreaks

\usepackage{mathrsfs} 

\usepackage{amsthm}
\newtheorem{theorem}{Theorem}
\newtheorem{prop}{Proposition}
\newtheorem{lemma}[theorem]{Lemma}
\newtheorem{definition}{Definition}
\newtheorem*{remark}{Remark}

\begin{document}
\title{Strong Asymptotic Composition Theorems for Sibson Mutual Information}


\author{Benjamin Wu\\
Electrical and Computer Engineering\\
Cornell University\\
Ithaca, NY 14850\\
\texttt{bhw49@cornell.edu}\\
\And
Aaron B. Wagner\\
Electrical and Computer Engineering\\
Cornell University\\
Ithaca, NY 14850\\
\texttt{wagner@cornell.edu}\\
\And
Ibrahim Issa\\
Dept. of Electrical and Computer Engineering\\
American University of Beirut\\
Beirut, Lebanon\\
\texttt{ii19@aub.edu.lb}\\
\And
G. Edward Suh\\
Electrical and Computer Engineering\\
Cornell University\\
Ithaca, NY 14850\\
\texttt{suh@ece.cornell.edu}\\
}

\maketitle

\begin{abstract}
    We characterize the growth of the Sibson and Arimoto mutual informations and $\alpha$-maximal leakage, 
of any order that is at least unity, between a random variable and a
growing set of noisy, conditionally independent and identically-distributed
observations of the random variable.
Each of these measures increases exponentially fast to a limit that is 
order- and measure-dependent, with an exponent that is order- and measure-independent.

\end{abstract}
\section{Introduction}

In the context of information leakage, composition theorems
characterize how leakage increases as a result of multiple,
independent, noisy observations of the sensitive data. 
Equivalently, they characterize how security (or privacy)
degrades under the ``composition'' of multiple observations
(or queries). In practice, attacks are often sequential in
nature, whether the application is side channels in computer
security~\cite{kocher_timing,wampler_information_2015,zhu_privacy_2012} or database privacy~\cite{Kairouz:Composition,
    Dwork:Concentrated, Mironov:Renyi}. Thus
composition theorems are practically relevant. They also
raise theoretical questions that are interesting in their
own right.

Various composition theorems for differential privacy and its variants
have been established (e.g.,~\cite{Kairouz:Composition,Dwork:Concentrated,
Mironov:Renyi}). For the
information-theoretic metrics of mutual information and
maximal leakage~\cite{Issa:ML,Issa:Shannon:ISIT,Issa:Metrics:ISIT,
Issa:ML:CISS} (throughout we assume discrete alphabets and base-2 logarithms)
\begin{align}
    \label{eq:MIdef}
    I(X;Y) & = \sum_{x,y} P(x,y) \log \frac{P(x,y)}{P(x) P(y)} \\
    \label{eq:MLdef}
    \mathcal{L}(X \rightarrow Y) & = \log
    \sum_{y} \max_{x: P(x) > 0} P(y|x),
\end{align}
and $\alpha$-maximal leakage~\cite{liao_tunable_2018},
less is known. While some results are available in the case that $P(y|x)$ is 
not known~\cite{smith_tight_2017}, here we assume it is known. For the metrics in 
(\ref{eq:MIdef})-(\ref{eq:MLdef}) it is straightforward to
show the ``weak'' composition theorem that if $Y_1,\ldots,Y_n$
are conditionally independent given $X$, then
\begin{align*}
I(X;Y^n) & \le \sum_{i = 1}^n I(X;Y_i) \\
\mathcal{L}(X \rightarrow Y^n) & \le \sum_{i = 1}^n 
           \mathcal{L}(X \rightarrow Y_i).
\end{align*}
These bounds are indeed weak in that if $Y_1,\ldots,Y_n$
are conditionally i.i.d.\ given $X$, then as $n \rightarrow \infty$,
the right-hand sides generally tend to infinity while the left-hand sides
remain bounded. A ``strong'' (asymptotic) composition theorem would 
identify the limit and characterize the speed of convergence.

We prove such a result for both mutual information and maximal
leakage. The limits are readily identified as the entropy and
$\log$-support size, respectively, of a minimal sufficient statistic of
$Y$ given $X$. In both cases, the speed of convergence to the
limit is exponential, and the exponent is the same.
Specifically, it is the minimum Chernoff information among
all pairs of distinct distributions $Q_{Y|X}(\cdot|x)$ and $Q_{Y|X}(\cdot|x')$.

Mutual information and maximal leakage are both instances of
Sibson mutual information~\cite{Sibson:MI,Verdu:Sibson,Issa:ML:CISS}, the 
former being order $1$ and the 
latter being order $\infty$. The striking fact that the
exponents governing the convergence to the limit
are the same at these two extreme points suggests that
Sibson mutual information of all orders satisfies a strong
asymptotic composition theorem, with the convergence rate
(but not the limit) being independent of the order. 
Meanwhile, Shannon mutual information can also be 
viewed as Arimoto mutual information of order $1$~\cite{arimoto_MI_1975}, 
and $\alpha$-maximal leakage is equivalently expressed as a maximization of 
Sibson or Arimoto mutual information of order $\alpha$
over $P(X)$ for $\alpha > 1$; for $\alpha = 1$, it equals Shannon 
mutual information~\cite{liao_tunable_2018}, as opposed
to the Shannon capacity.
Due to the intimate interrelation between these measures,
it is reasonable to suspect that similar strong asymptotic 
composition theorems obtain for them all.
Indeed, we prove strong composition theorems for Sibson
mutual information, Arimoto mutual information, and 
$\alpha$-maximal leakage, for all orders of at least unity.
In particular, we find that they all approach their respective
limits at the same $\alpha$-independent exponential rate, namely the 
minimum Chernoff information mentioned earlier.

The composition theorems proven here are different in
nature from those in the differential privacy 
literature.
Here we assume that the relevant probability distributions
are known, and we characterize the growth of leakage with repeated
looks from those distributions. We also assume
that $Y_1,\ldots,Y_n$ are conditionally i.i.d.\
given $X$. Composition theorems in differential privacy consider
the worst-case distributions given leakage levels for each
of $Y_1,\ldots,Y_n$ individually, assuming only conditional
independence.

Although our motivation is averaging attacks
in side channels, the results may have some use in capacity
studies of channels with multiple conditionally i.i.d.\ outputs
given the input~\cite[Prob.~7.20]{cover_elements_2006}.

%
%
%
%
%

\section{Sibson, Arimoto, R\'{e}nyi, and Chernoff}

This study relies on both Sibson's and Arimoto's tunable mutual information metrics as well as $\alpha$-maximal leakage. All random variables
in the paper are assumed discrete.

\begin{definition}[\hspace{1sp}\cite{Sibson:MI,Verdu:Sibson}]
The \emph{Sibson mutual information of order
$\alpha$} between random variables $X$ and $Y$ is defined by 
    \begin{equation}
        \label{eq:def:sibson}
        I_\alpha^S(X;Y)=\frac{\alpha}{\alpha-1}\log \sum_{y\in\mathcal{Y}}\Big( \sum_{x\in\mathcal{X}} P(x)P(y|x)^\alpha \Big)^{1/\alpha},
    \end{equation}
for $\alpha\in(0,1)\cap(1,\infty)$ and for $\alpha=1$ and $\alpha=\infty$ by its continuous extensions. These are
\begin{align*}
    I_1^S(X;Y) & = I(X;Y) \\
    I_\infty^S(X;Y) & = \mathcal{L}(X\rightarrow Y),
\end{align*}
defined in~(\ref{eq:MIdef})-(\ref{eq:MLdef}) above.
\end{definition}

\begin{definition}[\hspace{1sp}\cite{arimoto_MI_1975}]
The \emph{Arimoto mutual information of order
$\alpha$} between random variables $X$ and $Y$ is defined by 
    \begin{equation}
        \label{eq:def:arimoto}
        I_\alpha^A(X;Y)=\frac{\alpha}{\alpha-1}\log \sum_{y\in\mathcal{Y}}\Big( \frac{\sum_{x\in\mathcal{X}} P(x)^\alpha P(y|x)^\alpha}{\sum_{x\in\mathcal{X}}P(x)^\alpha  } \Big)^{1/\alpha}
    \end{equation}
for $\alpha\in(0,1)\cap(1,\infty)$ and for $\alpha=1$ and $\alpha=\infty$ by its continuous extensions. Note that~\cite{arimoto_MI_1975}
\begin{align*}
    I_1^A(X;Y) & = I(X;Y)
\end{align*}
but
\begin{align*}
    I_\infty^A(X;Y) & \neq \mathcal{L}(X\rightarrow Y).
\end{align*}
\end{definition}

\begin{definition}[\hspace{1sp}\cite{liao_tunable_2018}]
    The \emph{$\alpha$-maximal leakage} for $\alpha\in (1,\infty]$ is equivalently defined using either Sibson or Arimoto mutual information as:\footnote{The second equality for $1 < \alpha < \infty$ in (\ref{eq:maxdef}) is apparent from 
    (\ref{eq:def:sibson}) and (\ref{eq:def:arimoto})
    since the tilting of $P(x)$ in the latter can be absorbed into the maximization.}
\begin{equation}
    \label{eq:maxdef}
    \mathcal{L}_\alpha^{max}(X\rightarrow Y) = \max_{Q(X)}I_\alpha^S(X;Y) = \max_{Q(X)}I_\alpha^A(X;Y),
\end{equation}
where the maxima are over all distributions of
$X$ that have full support. 
For $\alpha=1$, we have
\begin{equation}
    \label{eq:maxdef2}
    \mathcal{L}_\alpha^{max}(X\rightarrow Y) = I(X;Y).
\end{equation}
    as opposed to the (Shannon) capacity
\begin{equation}
    \label{eq:capdef}
    \mathcal{C}(X;Y) = \max_{Q(X)} I(X;Y).
\end{equation}
\end{definition}

Liao \emph{et al.}~\cite{liao_tunable_2018} define $\alpha$-maximal
leakage operationally. The identities 
in~(\ref{eq:maxdef})-(\ref{eq:maxdef2}) are a theorem
in that work, which we shall take as a definition.
Likewise, Issa \emph{et al.}~\cite{Issa:ML} define maximal 
leakage operationally, and (\ref{eq:MLdef}) is a theorem that
we take as a definition.

We are interested in how $I_\alpha^S(X;Y^n)$, $I_\alpha^A(X;Y^n)$, and $\mathcal{L}_\alpha^{max}(X\rightarrow Y^n)$ grow with $n$ when
$Y_1,\ldots,Y_n$ are conditionally i.i.d.\ given $X$ for $\alpha \ge 1$.
The question for $\alpha < 1$ is meaningful in all cases but is not considered here because we are interested in the behavior of operational 
leakage measures, and the $\alpha < 1$ regime is not known
to be relevant to measuring leakage. We do not consider
the mutual information meaures put forward by 
Csisz\'{a}r~\cite{Csiszar:Cutoff} and Lapidoth and 
Pfister~\cite{Lapidoth:Two,Lapidoth:Testing:ITW} for the
same reason. For the quantities under study, 
we shall
see that the limits are given by \emph{R\'{e}nyi entropy}. As they will be needed for proof later, we also define \emph{Arimoto-R\'{e}nyi conditional entropy} and \emph{R\'{e}nyi divergence}.

\begin{definition}
    The \emph{R\'{e}nyi entropy} of order $\alpha$ of a random variable $X$ is given by:
    \begin{equation}
        H_\alpha(X) = \frac{1}{1-\alpha}\log\sum_{x\in\mathcal{X}}P(x)^\alpha
    \end{equation}
    for $\alpha\in(0,1)\cap(1,\infty)$ and for $\alpha=0$, $\alpha=1$, and $\alpha = \infty$ 
    by its continuous extensions. These are
\begin{align}
    H_0(X) & = \log |\{x: P(x) > 0\}| \\
    H_1(X) & = H(X) \\
    H_\infty(X) & = \log \frac{1}{\max_x P(x)}.
\end{align}
where $H(X)$ is the regular Shannon entropy.
\end{definition}

\begin{definition}
    The \emph{Arimoto-R\'{e}nyi conditional entropy} of order $\alpha$ of a random variable $X$ given $Y$ is defined as:
    \begin{equation}
        H_\alpha(X|Y) = \frac{\alpha}{1-\alpha}\log\sum_{y\in\mathcal{Y}}\Big( \sum_{x\in\mathcal{X}} P(x)^\alpha P(y|x)^\alpha \Big)^\frac{1}{\alpha}.
    \end{equation}
\end{definition}

\begin{remark}
    One can verify that it holds
    \begin{equation}
      I_\alpha^A(X;Y) = H_\alpha(X) - H_\alpha(X|Y).
    \end{equation}

\end{remark}

\begin{definition}
    The \emph{R\'{e}nyi divergence} of order $\alpha$ between probability distributions $P$ and $Q$ is defined for $\alpha\in[0,\infty)$, $\alpha \ne 1$ as:
    \begin{equation}
        D_\alpha(P||Q)=\frac{1}{\alpha-1}\log\sum_{x\in\mathcal{X}} P(x)^\alpha Q(x)^{1-\alpha},
    \end{equation}
    where the continuous extension at $\alpha=1$ is given by the standard Kullback-Leibler divergence 
    \begin{equation}
    D(P||Q) = \sum_{x \in \mathcal{X}} P(x) \log \frac{P(x)}{Q(x)}.
    \end{equation}

\end{definition}

The speed of convergence of $I_\alpha^S(X;Y^n)$, $I_\alpha^A(X;Y^n)$,  $\mathcal{L}_\alpha^{max}(X\rightarrow Y^n)$, and $\mathcal{C}{(X;Y^n)}$ and to their respective limits turns out to be governed by \emph{Chernoff information.}

\begin{definition}[\hspace{1sp}\cite{cover_elements_2006}]
    The \emph{Chernoff information} between two probability mass functions, $P_1$ and $P_2$, over the same alphabet $\mathcal{X}$ is given as follows. First, for all $x\in\mathcal{X}$ and $\lambda\in[0,1]$, let:
    \begin{equation}
        P_\lambda(x)=P_\lambda(P_1,P_2,x)=\frac{P_1(x)^\lambda P_2(x)^{1-\lambda}}{\sum_{x^\prime\in\mathcal{X}}P_1(x^\prime)^\lambda P_2(x^\prime)^{1-\lambda}}.
        \label{eq:tilteddiv}
    \end{equation}
    Then the Chernoff information is given by
    \begin{equation}
        \mathscr{C}(P_1||P_2) = D(P_{\lambda^*}||P_1) = D(P_{\lambda^*}||P_2),
    \end{equation}
    where $\lambda^*$ is any value of $\lambda$ such that the above two relative entropies are equal. Equivalently, the Chernoff information is also given by:
    \begin{equation}
        \label{eq:Chernoff:alt}
        \mathscr{C}(P_1||P_2) = -\min_{0\leq\lambda<1}{\log\left(\sum_x P_1(x)^\lambda P_2(x)^{1-\lambda}\right)}\\
    \end{equation}
\end{definition}
Since we consider finite alphabets, the Chernoff information
is infinite if and only if $P_1$ and $P_2$ have
disjoint support.

\emph{Other Notation:}
We use $\mathcal{P}_n$ to denote the set of all possible
empirical distributions of $Y^n$.
We let $\mathcal{P}$ denote the set of all possible probability distributions 
over $\mathcal{Y}$
For any $P\in\mathcal{P}$, let
$$T(P)=\{y^n\in\mathcal{Y}^n|P_{y^n}=P\},$$
where $P_{y^n}$ is the empirical distribution of $y^n$. Note that $T(P)$ is empty if $P\notin\mathcal{P}_n$.
We use $Q(\cdot)$ to denote the true distributions of $X$ and $Y^n$.
We let $Q_x$ denote the distribution of $Y$ given $x$ for a given $x\in\mathcal{X}$. For any $P\in\mathcal{P}$, let $x_k(P)$ denote $x\in\mathcal{X}$ such that $D(P||Q_x)$ is the $k^{th}$ smallest relative entropy across all elements of $\mathcal{X}$. Ties can be broken by the ordering of $\mathcal{X}$.

We also define $x$-domains for fixed $n$ in two slightly different ways. Let

    \begin{equation}\label{xDomainDef}
    D_x = \{P\in\mathcal{P} | D(P||Q_x)<D(P||Q_{x^\prime}) \ \forall x^\prime\neq x\}
    \end{equation}
    \begin{equation}\label{xDomainDef2}
    \bar{D}_x = \{P\in\mathcal{P} | D(P||Q_x)\leq D(P||Q_{x^\prime}) \ \forall x^\prime\in\mathcal{X}\}
    \end{equation}


    Note that for any $P\in \bar{D}_x$, 
    $D(P||Q_x)=\min_{x^\prime\in\mathcal{X}}D(P||Q_{x^\prime})$. 

\section{The Result}
\label{sec:result}

Let $X$ be a random variable with alphabet $\mathcal{X}=\{x_1, x_2,...x_{|\mathcal{X}|}\}$. Let $Y^n=(Y_1, Y_2,...Y_n)$ be a vector of discrete random variables with a shared alphabet $\mathcal{Y}=\{y_1, y_2,...y_{|\mathcal{Y}|}\}$.
We assume that $Y_1, Y_2,\ldots,Y_n$ are conditionally i.i.d.\ given $X$.
We may assume, without loss of generality, that $X$ and $Y$ have
full support. We will also assume that 
the distributions $P_{Y|X}(\cdot|x)$ are unique over $x$, which 
we call the \emph{unique row assumption}. For Sibson mutual
information and $\alpha$-max leakage, this is without loss
of generality, since we can divide $\mathcal{X}$ into equivalence
classes based on their respective $P_{Y|X}(\cdot|x)$ distributions
and define $\tilde{X}$ to be the equivalence class of $X$. Then
both Markov chains $X \leftrightarrow \tilde{X} \leftrightarrow Y^n$ 
and $\tilde{X} \leftrightarrow X \leftrightarrow Y^n$ hold and so
\begin{align}
    I_\alpha^S(X;Y^n) & = I_\alpha^S(\tilde{X};Y^n) \\
    \mathcal{L}_\alpha^{max}(X\rightarrow Y^n) & = 
    \mathcal{L}_\alpha^{max}(\tilde{X}\rightarrow Y^n),
\end{align}
by the data processing inequality for Sibson 
mutual information~\cite{Polyanskiy:Allerton:Arimoto}
and $\alpha$-maximal leakage~\cite[Thm.~3]{liao_tunable_2018}.
We may then replace $X$ with $\tilde{X}$ in
the case of these measures. For Arimoto mutual
information, the chain rule does not hold, and
in fact an arbitrarily large discrepancy can exist 
between
$I_\alpha^A(X;Y)$ and $I_\alpha^A(\tilde{X};Y)$,
as shown in Appendix~\ref{app:arimotocounter}, where
it is also shown that the unique row assumption is 
nonetheless still without loss of generality.

Our measures of interest satisfy the following
upper bounds:
\begin{align}
    \label{eq:MIbound}
    I(X;Y^n) & \le H(X) \\
    \label{eq:capbound}
    \mathcal{C}(X;Y^n) & \le \log|\mathcal{X}| \\
    \label{eq:sibsonbound}
    I_\alpha^S(X;Y^n) & \le H_{1/\alpha}(X) \ \ \mbox{\cite[Ex.~2 and Thm.~3]{Verdu:Sibson}} \\
    \label{eq:arimotobound}
    I_\alpha^A(X;Y^n) & \le H_{\alpha}(X) \ \ \mbox{\cite[Prop.~3]{Fehr:Arimoto}} \\
    \label{eq:alphamaxbound}
    \mathcal{L}_\alpha^{max}(X\rightarrow Y^n) & \le 
    \begin{cases}
        H(X) & \text{if $\alpha = 1$} \\
        \log{|\mathcal{X}|} & \text{if $\alpha > 1$}
    \end{cases}
        \mbox{\cite[Thm.~3]{liao_tunable_2018}} \\
        \nonumber
        & =: \mathcal{L}_\alpha(X),
\end{align}
where each inequality holds for all $n$ and all $\alpha \in [1,\infty]$.  
Comparing~(\ref{eq:sibsonbound}) and~(\ref{eq:arimotobound})
suggests that perhaps the Arimoto mutual
information of order $\alpha$ should be associated with
the Sibson mutual information of order $1/\alpha$;
the identity in (\ref{eq:maxdef}) suggests otherwise.

Our main result describes how fast 
these upper bounds are approached as $n \rightarrow \infty$.
\begin{theorem}
Under the unique row assumption, for all $\alpha \in [1,\infty]$,
\begin{align} 
    \min_{x \ne x^\prime} \mathscr{C}(Q_x||Q_{x^\prime}) & = 
    \label{eq:MIresult}
    \lim_{n\rightarrow \infty}-\frac{1}{n}\log\Big(H(X) - I(X;Y^n) \Big)  \\
    \label{eq:capresult}
    & = \lim_{n\rightarrow \infty}-\frac{1}{n}\log\Big(\log|\mathcal{X}| - \mathcal{C}(X;Y^n) \Big)  \\
    \label{eq:Sibsonresult}
    & = \lim_{n\rightarrow \infty}-\frac{1}{n}\log\Big( H_{1/\alpha}(X) - I_\alpha^S(X;Y^n) \Big)  \\
    \label{eq:Arimotoresult}
   & = \lim_{n\rightarrow \infty}-\frac{1}{n}\log\Big( H_{\alpha}(X) - I_\alpha^A(X;Y^n) \Big)  \\
    \label{eq:maxalpharesult}
    & = \lim_{n\rightarrow\infty}-\frac{1}{n}\log\Big(\mathcal{L}_\alpha(X) - \mathcal{L}_\alpha^{max}(X\rightarrow Y^n)\Big).
\end{align}
    \label{thm:main}
\end{theorem}
Thus the Chernoff information governs the exponential 
rate-of-approach for all measures and for all 
values of $\alpha$.  This Chernoff information is
infinite if $Q_x$ and $Q_{x^\prime}$ have disjoint
support for all $x \ne x^\prime$; in this case,
the bounds in (\ref{eq:MIbound})-(\ref{eq:alphamaxbound})
are met with equality already for $n = 1$. Channels
with this property arise naturally in certain
applications~\cite{Wu:CNS}.

Observe that (\ref{eq:Sibsonresult})-(\ref{eq:maxalpharesult}) coincide 
with (\ref{eq:MIresult}) when $\alpha = 1$. Also, (\ref{eq:Sibsonresult}) and
(\ref{eq:maxalpharesult}) coincide for $\alpha = \infty$;
otherwise the assertions are independent.

For continuous random variables, it is meaningful and
interesting to study how $I_\alpha^S(X;Y^n)$,
$\mathcal{C}(X;Y^n)$, and 
$\mathcal{L}_\alpha^{max}(X\rightarrow Y^n)$ grow
with $n$. The behavior would be fundamentally
different from the discrete case, however. 
See Aishwarya and Madiman~\cite{Aishwarya:Arimoto} for 
a discussion of Arimoto mutual information in
the continuous case.

The remainder of the paper is 
devoted to proving the various assertions
contained within Theorem~\ref{thm:main}. The assertions
are evidently asymptotic in nature, and our proofs
are not opimtized to provide the best finite-$n$
bounds. Numerical experiments show that in many cases
our lower and upper bounds are quite far apart for
moderate values of $n$.

\section{Proof for Mutual Information and Capacity}
    We begin by proving (\ref{eq:MIresult}) and (\ref{eq:capresult}), starting
    with the former.
    For this, we derive separate upper and lower bounds 
    on $-H(X|Y^n)$.  For the lower bound,
	\begin{align}
	&-H(X|Y^n) \equiv \sum_{y^n\in\mathcal{Y}^n} Q(y^n) \sum_{x\in\mathcal{X}}  Q(x|y^n)\log Q(x|y^n)\\
	&= \sum_{P\in\mathcal{P}_n} \sum_{y^n\in T(P)} Q(y^n) \sum_{x\in\mathcal{X}}  \frac{Q(y^n|x)Q(x)}{Q(y^n)}\log \frac{Q(y^n|x)Q(x)}{Q(y^n)}\\
	&= \sum_{P\in\mathcal{P}_n} \sum_{y^n\in T(P)} \sum_{x\in\mathcal{X}}  \frac{1}{|T(P)|} Q(T(P)|x)Q(x)\notag\\
	&\phantom{====}\cdot\log \frac{\frac{1}{|T(P)|}Q(T(P)|x)Q(x)}{ \sum_{x^\prime\in\mathcal{X}} \frac{1}{|T(P)|}Q(T(P)|x^\prime)Q(x^\prime)}\\
	&= \sum_{P\in\mathcal{P}_n}  \sum_{x\in\mathcal{X}}  Q(T(P)|x)Q(x)\log \frac{Q(T(P)|x)Q(x)}{ \sum_{x^\prime\in\mathcal{X}} Q(T(P)|x^\prime)Q(x^\prime)}\\
	&= - \sum_{\substack{P\in\mathcal{P}_n: \\ Q(T(P))>0}}\bigg[  Q(T(P)|x_1(P))Q(x_1(P))\notag\\
	&\phantom{====}\cdot\log \frac{ \sum_{x^\prime\in\mathcal{X}} Q(T(P)|x^\prime)Q(x^\prime)}{Q(T(P)|x_1(P))Q(x_1(P))} \notag\\ 
	& \phantom{==}+ \sum_{\substack{x\neq x_1(P): \\ Q(T(P)|x)>0}}  Q(T(P)|x)Q(x)\notag\\
	&\phantom{====}\cdot\log \frac{ \sum_{x^\prime\in\mathcal{X}} Q(T(P)|x^\prime)Q(x^\prime)}{Q(T(P)|x)Q(x)}      \bigg],
	\end{align}
	due to the convention that  $0\log 0=0$. Then, replacing weighted sums over $x$ with their largest summand gives
	\begin{align}
	&\geq - \sum_{\substack{P\in\mathcal{P}_n: \\ Q(T(P))>0}}\bigg[  Q(T(P)|x_1(P))Q(x_1(P))\notag\\
	&\phantom{====}\cdot\log \Big(1+ \frac{ \sum_{x^\prime\neq x_1(P)} Q(T(P)|x^\prime)Q(x^\prime)}{Q(T(P)|x_1(P))Q(x_1(P))}\Big) \notag\\ 
	& \phantom{==}+ \max_{\substack{x\neq x_1(P): \\ Q(T(P)|x)>0}} \Big\{ Q(T(P)|x)\log \frac{ \max_{x^\prime\in\mathcal{X}} Q(T(P)|x^\prime)}{Q(T(P)|x)Q(x)}\Big\} \bigg].
	\end{align}
	Note that the entire expression inside the summation over $P$ is 0 if $Q(T(P)|x_2(P))=0$. Letting $Q_{\min}(X)=\min_{x\in\mathcal{X}}Q(x)$ and using $\ln(1+x)\leq x$ for the $x=x_1(P)$ term,
	\begin{align}
	&\geq - \sum_{\substack{P\in\mathcal{P}_n: \\ Q(T(P))>0}}\bigg[  \frac{1}{\ln 2}\sum_{x^\prime\neq x_1(P)} Q(T(P)|x^\prime)Q(x^\prime) \notag\\ 
 & \phantom{==}+ \max_{\substack{x\neq x_1(P): \\ Q(T(P)|x)>0}} \Big\{ Q(T(P)|x)  \Big\} \notag \\
 & \phantom{====} \cdot \log \frac{1}{\min_{\substack{x\neq x_1(P): \\ Q(T(P)|x)>0}}Q(T(P)|x)\cdot Q_{\min}(X)} \bigg]\\
	&\geq - \sum_{\substack{P\in\mathcal{P}_n: \\ Q(T(P))>0}}\bigg[  \frac{1}{\ln 2}2^{-nD(P||Q_{x_2(P)})}  + 2^{-nD(P||Q_{x_2(P)})}\notag\\
	&\phantom{====} \cdot\big[nD_{sup}+\log \frac{(n+1)^{|\mathcal{Y}|}}{Q_{\min}(X)}\big] \bigg]
\end{align}
where
\begin{align}
	&D_{sup}\equiv\sup_{\substack{x, P^\prime\in\mathcal{P} \\ D(P^\prime||Q_x)<\infty}} D(P^\prime||Q_x)\\
	&=\sup_{\substack{x, P^\prime\in\mathcal{P}: \\ D(P^\prime||Q_x)<\infty}} \sum_{y\in\mathcal{Y}}P^\prime(y)\log{\frac{P^\prime(y)}{Q(y|x)}}\\
	&=\sup_{\substack{x, P^\prime\in\mathcal{P}: \\ D(P^\prime||Q_x)<\infty}} \sum_{y\in\mathcal{Y}}P^\prime(y)\log{\frac{1}{Q(y|x)}} - H(P^\prime)\\
	&\leq \sup_{x}\log{\frac{1}{\min_{Q(y|x)>0}Q(y|x)}}< \infty.
\end{align}
Hence,
\begin{align}
	&-H(X|Y^n) \notag\\
	&\geq -(n+1)^{|\mathcal{Y}|} 2^{-nD_n^*}\big[   \frac{1}{\ln 2}  +  \log \frac{ (n+1)^{|\mathcal{Y}|} }{Q_{\min}(X)} + nD_{sup}      \big] \label{condEntrUpper}
\end{align}
where
\begin{equation}\label{D_n^*}
	D_n^*=\min_{P\in\mathcal{P}_n}D(P||Q_{x_2(P)})
\end{equation}
and $P_n^*$ is its minimizer.

For the upper bound,
\begin{align}
	&-H(X|Y^n)\notag\\
	&=\sum_{P\in\mathcal{P}_n}  \sum_{x\in\mathcal{X}}  Q(T(P)|x)Q(x)\log \frac{Q(T(P)|x)Q(x)}{ \sum_{x^\prime\in\mathcal{X}} Q(T(P)|x^\prime)Q(x^\prime)}\\
	&\leq \sum_{x\in\mathcal{X}}  Q(T(P_n^*)|x)Q(x)\log \frac{Q(T(P_n^*)|x)Q(x)}{ \sum_{x^\prime\in\mathcal{X}} Q(T(P_n^*)|x^\prime)Q(x^\prime)}\\
	&\leq  Q(T(P_n^*)|x_1(P_n^*))Q(x_1(P_n^*))\notag\\
	&\phantom{====}\cdot\log \frac{Q(T(P_n^*)|x_1(P_n^*))Q(x_1(P_n^*))}{ \sum_{x^\prime\in\mathcal{X}} Q(T(P_n^*)|x^\prime)Q(x^\prime)}\\
	&=  Q(T(P_n^*)|x_1(P_n^*))Q(x_1(P_n^*))\notag\\
	&\phantom{====}\cdot\log\big[1- \frac{\sum_{x^\prime\neq x_1(P_n^*)} Q(T(P_n^*)|x^\prime)Q(x^\prime)}{ \sum_{x^\prime\in\mathcal{X}} Q(T(P_n^*)|x^\prime)Q(x^\prime)}\big]\\
    \intertext{recalling that $-\ln(1-x)\geq x$,}
	&\leq - Q(T(P_n^*)|x_1(P_n^*))Q(x_1(P_n^*))\notag\\
	&\phantom{====}\cdot\frac{\sum_{x^\prime\neq x_1(P_n^*)} Q(T(P_n^*)|x^\prime)Q(x^\prime)}{ \sum_{x^\prime\in\mathcal{X}} Q(T(P_n^*)|x^\prime)Q(x^\prime)}\cdot\frac{1}{\ln 2}
	\end{align}
	\begin{align}
	&\leq -  Q(T(P_n^*)|x_1(P_n^*))Q(x_1(P_n^*))\notag\\
	&\phantom{====}\cdot\frac{Q(T(P_n^*)|x_2(P_n^*))Q(x_2(P_n^*))}{ \max_{x^\prime\in\mathcal{X}}Q(T(P_n^*)|x^\prime)}\cdot\frac{1}{\ln 2}\\
	&\leq -  \frac{1}{(n+1)^{|\mathcal{Y}|}}2^{-nD(P_n^*||Q_{x_1(P_n^*)})}Q(x_1(P_n^*))\notag\\
	&\phantom{====}\cdot\frac{2^{-nD_n^*}Q(x_2(P_n^*))}{ (n+1)^{|\mathcal{Y}|}2^{-nD(P_n^*||Q_{x_1(P_n^*)})}}\cdot\frac{1}{\ln 2}\\
	&= -  \frac{Q(x_1(P_n^*))Q(x_2(P_n^*))}{(n+1)^{2|\mathcal{Y}|}\ln{2}}2^{-nD_n^*}.\label{mutualInf_upper}
\end{align}

As we have now shown that mutual information is upper and lower bounded by expressions of the form $H(X)-K_n\cdot 2^{-nD_n^*}$ for some subexponential sequence $K_n$, it remains to be shown that this exponent approaches the minimum Chernoff information as $n\rightarrow\infty$.

First, it can be shown using standard continuity arguments that
\begin{equation}
	\lim_{n\rightarrow\infty}\inf_{P\in\mathcal{P}_n} D(P||Q_{x_2(P)})=\inf_{P\in\mathcal{P}} D(P||Q_{x_2(P)})
\end{equation}
since $D(P||Q_{x_2(P)})$ is a continuous function of $P$. Finally, we arrive at the desired result using Lemma \ref{ChernoffLemma_journal} in Appendix~\ref{app:ancillary}.

Turning to the result for capacity, let $Q_u$ denote the uniform distribution
over $\mathcal{X}$. Then by~(\ref{eq:MIresult}) we have
\begin{align}
    &    \liminf_{n \rightarrow \infty} - \frac{1}{n} \log \left( \log|\mathcal{X}| -
                      C(X;Y^n)\right) \\
                       & \ge
    \liminf_{n \rightarrow \infty} - \frac{1}{n} \log \left( \log|\mathcal{X}| -
                       I(X;Y^n)\Big|_{Q_u} \right) \\
                      & = \min_{x \ne x^\prime}
                          \mathscr{C}(Q_x||Q_{x'}).
\end{align}
For the reverse inequality, for each $n$, let $Q_n$ be a maximizer
of $I(X;Y^n)$. Then from the previous observation, eventually we have
\begin{align}
    H(X)\Big|_{Q_n} -  H(X|Y^n)\Big|_{Q_n} & \ge
     I(X;Y^n)\Big|_{Q_u} \\
         & \ge \log|\mathcal{X}| - e^{-\frac{n}{2} \min_{x \ne x^\prime}
             \mathscr{C}(Q_x||Q_{x^\prime})}.
\end{align}
Dropping the second term from the left-hand side and using the fact
that 
\begin{equation}
    D(Q_n||Q_u) = \log|\mathcal{X}| - H(X)\Big|_{Q_n} 
\end{equation}
this implies that, eventually,
\begin{equation}
         e^{-\frac{n}{2} \min_{x \ne x^\prime}
             \mathscr{C}(Q_x||Q_{x^\prime})} \ge D(Q_n||Q_u).
\end{equation}
Thus $Q_n$ tends to $Q_u$ as $n \rightarrow \infty$. Combining this
fact with the bound in~(\ref{mutualInf_upper}), we have that, eventually,
\begin{align}
    \mathcal{C}(X;Y^n) & = 
    I(X;Y^n)\Big|_{Q_n} \\
    & = H(X)\Big|_{Q_n} -  H(X|Y)\Big|_{Q_n} \\
         & \le H(X)\Big|_{Q_n} -  \frac{Q_n(x_1(P_n^*))Q(x_2(P_n^*))}{(n+1)^{2|\mathcal{Y}|}
                 \ln 2} 2^{-n D_n^*}, \\
         & \le H(X)\Big|_{Q_n} -  \frac{1}{4 |\mathcal{X}|^2 (n+1)^{2|\mathcal{Y}|}
                 \ln 2} 2^{-n D_n^*} \\
         & \le \log |\mathcal{X}| -  \frac{1}{4 |\mathcal{X}|^2 (n+1)^{2|\mathcal{Y}|}
                 \ln 2} 2^{-n D_n^*},
\end{align}
which establishes the result since $D^*_n$ converges to the Chernoff information
as shown above.

\section{Proof for Sibson ($\alpha\in(1,\infty)$)}\label{sec::sibsonproof}
We turn to~(\ref{eq:Sibsonresult}), focusing on the regime
$\alpha\in(1,\infty)$, since the $\alpha = 1$ case is 
    established in (\ref{eq:MIresult}) and the $\alpha = \infty$ case
    will be proven subsequently.
First, we derive a lower bound of $I_\alpha^S(X;Y^n)$ for $\alpha>1$
that will be useful in this and subsequent proofs.
\begin{lemma}\label{lowerboth}
    \begin{equation}\label{eq:lowerboth}
        I_\alpha^S(X;Y^n) \geq H_{1/\alpha}(X)- \frac{\alpha}{(\alpha-1)\ln 2}\left(\Gamma_n+\frac{\Gamma_n^2}{2(1-\Gamma_n)}\right)
    \end{equation}
    for $\alpha>1$, where
    \begin{align}
        \Gamma_n & = \min(1,(n+1)^{|\mathcal{Y}|}\cdot 2^{-n \cdot \min_{x \ne x'} \mathscr{C}(Q_x||Q_{x'})}).\notag\\
    \end{align}
\end{lemma}

\begin{remark}
    If $Q_x$ and $Q_{x^\prime}$ have disjoint support for every
    $x \ne x^\prime$, then $\Gamma_n = 0$ and this lemma establishes that 
    $I_\alpha^S(X;Y^n) = H_{1/\alpha}(X)$ for any $n \ge 1$.
\end{remark}

\begin{proof}
We use the $D_x$ sets defined 
in (\ref{xDomainDef}) and (\ref{xDomainDef2}):
    \begin{align}
        &I_\alpha^S(X;Y^n)\equiv\frac{\alpha}{\alpha-1}\log \sum_{y^n\in\mathcal{Y}^n}\Big( \sum_{x\in\mathcal{X}} Q(x)Q(y^n|x)^\alpha \Big)^{1/\alpha}\\
        &=\frac{\alpha}{\alpha-1}\log \sum_{P\in\mathcal{P}_n}\Big( \sum_{x\in\mathcal{X}} Q(x)Q(T(P)|x)^\alpha \Big)^{1/\alpha}\\
        &\geq\frac{\alpha}{\alpha-1}\log \sum_{x\in\mathcal{X}}\sum_{P\in D_x\cap\mathcal{P}_n}\Big( \sum_{x^\prime\in\mathcal{X}} Q(x^\prime)Q(T(P)|x^\prime)^\alpha \Big)^{1/\alpha}\\
        &\geq \frac{\alpha}{\alpha-1}\log \sum_{x\in\mathcal{X}} Q(x)^{1/\alpha}\sum_{P\in D_x\cap\mathcal{P}_n}Q(T(P)|x)\\
        &= \frac{\alpha}{\alpha-1}\log \sum_{x\in\mathcal{X}} Q(x)^{1/\alpha}\Big(1-\sum_{P\in \mathcal{P}_n\backslash D_x}Q(T(P)|x)\Big)\\
        &= \frac{\alpha}{\alpha-1}\log \Big(\sum_{x\in\mathcal{X}} Q(x)^{1/\alpha} \\
        \nonumber
          & \phantom{= \frac{\alpha}{1 - \alpha}} - \sum_{x\in\mathcal{X}}\sum_{P\in \mathcal{P}_n\backslash D_x}Q(x)^{1/\alpha}Q(T(P)|x)\Big).
    \end{align}
    Define 
    \begin{equation}
       \gamma_n = \frac{\sum_{x\in\mathcal{X}}\sum_{P\in \mathcal{P}_n\backslash D_x}Q(x)^{1/\alpha}Q(T(P)|x)}{\sum_{x\in\mathcal{X}} Q(x)^{1/\alpha}} \le 1.
    \end{equation}
    Then we can write
    \begin{align}
        I_\alpha^S(X;Y^n) & \geq \frac{\alpha}{\alpha-1}\log \Big\{\Big(\sum_{x\in\mathcal{X}} Q(x)^{1/\alpha}\Big)(1-\gamma_n)\Big\}\\
        &= H_{1/\alpha}(X)+\frac{\alpha}{\alpha-1}\log(1-\gamma_n).
    \end{align}
    Note that 
    \begin{align}
        &\ln(1-\epsilon) = -\sum_{i=1}^{\infty}{\frac{\epsilon^i}{i}}\\
        &\geq-\epsilon-\frac{\epsilon}{2}\Big(\sum_{i=1}^{\infty}{\epsilon^i}\Big)=-\epsilon-\frac{\epsilon^2}{2(1-\epsilon)} \label{natLogLowerBnd}
    \end{align}
    for $0<\epsilon<1$. Hence,
    \begin{equation} 
        I_\alpha^S(X;Y^n) \geq H_{1/\alpha}(X)+\frac{\alpha}{(\alpha-1)\ln 2} (-\gamma_n - \frac{\gamma_n^2}{2(1-\gamma_n)}).
    \end{equation}
    The right-hand side in decreasing in $\gamma_n$ over $[0,1]$. We also have
    \begin{align}
        &\gamma_n \leq \frac{\sum_{x\in\mathcal{X}}Q(x)^{1/\alpha}(n+1)^{|\mathcal{Y}|}\cdot\max_{P\in \mathcal{P}_n\backslash D_x}Q(T(P)|x)}{\sum_{x\in\mathcal{X}} Q(x)^{1/\alpha}}\\
        &\leq \frac{\sum_{x\in\mathcal{X}}Q(x)^{1/\alpha}(n+1)^{|\mathcal{Y}|}\cdot\underset{x^\prime\in\mathcal{X}}{\max}\underset{P\in \mathcal{P}_n\backslash D_{x^\prime}}{\max}Q(T(P)|x^\prime)}{\sum_{x\in\mathcal{X}} Q(x)^{1/\alpha}}\\
        &= (n+1)^{|\mathcal{Y}|}\cdot\max_{x\in\mathcal{X}}\max_{P\in \mathcal{P}_n\backslash D_x}Q(T(P)|x)\\
        &\leq (n+1)^{|\mathcal{Y}|}\cdot2^{-n(\min_{x\in\mathcal{X}}\min_{P\in \mathcal{P}_n\backslash D_x}D(P||Q_x))}\\
        &\leq (n+1)^{|\mathcal{Y}|}\cdot 2^{-n(\min_{x\neq x^\prime}\inf_{P\in \bar{D}_{x^\prime}}D(P||Q_x))}\\
        &= (n+1)^{|\mathcal{Y}|}\cdot 2^{-n \cdot \min_{x \ne x'} \mathscr{C}(Q_x||Q_{x'})},
    \end{align}
    where we have used Lemma~\ref{ChernoffLemma_journal} in Appendix~\ref{app:ancillary}.
\end{proof}

We next prove an analogous upper bound.
\begin{lemma}
    For $\alpha > 1$, define
    \begin{equation}
         F(x,P) = Q(x)Q(T(P)|x)^\alpha.
         \label{eq:Fdef}
    \end{equation}
    For each $n$, let $\{E_{x_i}^{(n)}\}_{i=1}^{|\mathcal{X}|}$ be a partition of $\mathcal{P}_n$ such that $P\in E_x^{(n)}$ implies 
                                        $F(x,P) = \max_{x^\prime\in\mathcal{X}}F(x^\prime,P)$. Then
    \begin{align}
        I_\alpha^S(X;Y^n) &\leq H_{1/\alpha}(X)+\frac{\alpha}{(\alpha-1)\ln 2}\frac{1}{\underset{x\in\mathcal{X}}{\sum}Q(x)^{1/\alpha}}\sum_{x\in\mathcal{X}}\sum_{P\not\in E_x^{(n)}}\notag\\
        &\phantom{====}\cdot(F(x_1(P),P)^{1/\alpha-1}-F(x,P)^{1/\alpha-1})F(x,P)),
    \end{align}
    where for the remainder of this section we redefine $x_k(P)$ so that they are ordered by $F(x,P)$ instead of relative entropy. Note that this ordering now depends on $n$.
    \label{upperboth}
\end{lemma}

\begin{proof}
    We have
    \begin{align}
        &I_\alpha^S(X;Y^n) = \frac{\alpha}{\alpha-1}\log \sum_{x\in\mathcal{X}}\sum_{P\in E_x^{(n)}}\Big( \sum_{x^\prime\in\mathcal{X}} F(x^\prime,P) \Big)^{1/\alpha}\\ \label{SibsonRetraceStart}
        &=\frac{\alpha}{\alpha-1}\log \sum_{x\in\mathcal{X}}\sum_{P\in E_x^{(n)}} F(x,P)^{1/\alpha}\Big( 1+\sum_{x^\prime\neq x} \frac{F(x^\prime,P)}{F(x,P)} \Big)^{1/\alpha}
    \end{align}
    Using the Taylor series expansion of $(1+x)^{1/\alpha}$ and discarding $x^2$ and higher-order terms (since $\frac{1}{\alpha}<1$), we have
    \begin{align}
        &\leq\frac{\alpha}{\alpha-1}\log \sum_{x\in\mathcal{X}}\sum_{P\in E_x^{(n)}}F(x,P)^{1/\alpha}\Big( 1+\frac{1}{\alpha}\sum_{x^\prime\neq x} \frac{F(x^\prime,P)}{F(x,P)} \Big)\\
        &\leq\frac{\alpha}{\alpha-1}\log \sum_{x\in\mathcal{X}}\sum_{P\in E_x^{(n)}} \Big( F(x,P)^{1/\alpha}\notag\\ 
        &\phantom{==}+ F(x,P)^{1/\alpha-1}\sum_{x^\prime\neq x} F(x^\prime,P)\Big),
    \end{align}
    where we have used the fact that $\alpha > 1$. Continuing,
    \begin{align}
        &=\frac{\alpha}{\alpha-1}\log \sum_{x\in\mathcal{X}}\Big(\sum_{P\in E_x^{(n)}} F(x,P)^{1/\alpha}\notag\\ 
        &\phantom{====}+ \sum_{P\not\in E_x^{(n)}}F(x_1(P),P)^{1/\alpha-1} F(x,P)\Big)\\
        &=\frac{\alpha}{\alpha-1}\log \sum_{x\in\mathcal{X}}\Big(\sum_{P\in E_x^{(n)}} F(x,P)^{1/\alpha}\notag\\ 
        &\phantom{====}+ \sum_{P\not\in E_x^{(n)}}F(x_1(P),P)^{1/\alpha-1} F(x,P)\notag\\
        &\phantom{====}+\sum_{P\not\in E_x^{(n)}}F(x,P)^{1/\alpha}-\sum_{P\not\in E_x^{(n)}}F(x,P)^{1/\alpha}\Big)\\
        &=\frac{\alpha}{\alpha-1}\log \sum_{x\in\mathcal{X}}\Big(\sum_{P\in\mathcal{P}_n} F(x,P)^{1/\alpha} \notag\\
        &\phantom{=}+ \sum_{P\not\in E_x^{(n)}}\big(F(x_1(P),P)^{1/\alpha-1}-F(x,P)^{1/\alpha-1}\big)F(x,P) \Big)
    \end{align}
    Using $\ln(1+x) \le x$ then gives the result.
\end{proof}

The lower bound in (\ref{eq:Sibsonresult}) for $\alpha \in (1,\infty)$ follows directly from Lemma~\ref{lowerboth}. 
For the upper bound, pick $x_a\neq x_b$ and $P^*\in D_{x_b}$. Let $\{P_n\}_{n=1}^\infty$ be a sequence of types converging to $P^*$. 
From Lemma~\ref{upperboth} we have
\begin{align}
    I_\alpha^S(X;Y^n)
        &\leq H_{1/\alpha}(X)+\frac{\alpha}{(\alpha-1)\ln 2}\frac{1}{\underset{x\in\mathcal{X}}{\sum}Q(x)^{1/\alpha}}\sum_{x\in\mathcal{X}}\sum_{P\not\in E_x^{(n)}}\notag\\
        &\phantom{====}\cdot(F(x_1(P),P)^{1/\alpha-1}-F(x,P)^{1/\alpha-1})F(x,P)).
\end{align}
Note that eventually $P_n\in E_{x_b}^{(n)}$, $x_1(P_n)=x_b$ and $F(x_b,P_n)^{1/\alpha-1}<\frac{1}{2}F(x_a,P_n)^{1/\alpha-1}$. Thus, eventually,
    \begin{align}
        &\leq H_{1/\alpha}(X)+\frac{\alpha}{(\alpha-1)\ln 2}\frac{1}{\underset{x\in\mathcal{X}}{\sum}Q(x)^{1/\alpha}}\notag\\
        &\phantom{=}\cdot(F(x_1(P_n),P_n)^{1/\alpha-1}-F(x_a,P_n)^{1/\alpha-1})F(x_a,P_n)).\label{SibsonRetraceEnd} \\
        &\leq H_{1/\alpha}(X)-\frac{\alpha}{2(\alpha-1)\ln 2}\frac{1}{\underset{x\in\mathcal{X}}{\sum}Q(x)^{1/\alpha}}F(x_a,P_n)^{1/\alpha}\\
        &\leq H_{1/\alpha}(X)-\frac{\alpha}{2(\alpha-1)\ln 2}\frac{1}{\underset{x \in \mathcal{X}}{\sum}Q(x)^{1/\alpha}}Q_{\min}(X)^{1/\alpha}\notag\\
        &\phantom{====} \cdot\frac{1}{(n+1)^{|\mathcal{Y}|}}2^{-nD(P_n||Q_{x_a})} 
        \label{eq:upperSibson}
    \end{align}
    where $Q_{\min}(X)=\min_{x\in\mathcal{X}}Q(x)$. This implies:
    \begin{align}
        &\underset{n\rightarrow\infty}{\lim\sup}-\frac{1}{n}\log\big( H_{1/\alpha}(X)-I_\alpha^S(X;Y^n) \big)\notag\\
        &\leq \lim_{n\rightarrow\infty}D(P_n||Q_{x_a})=D(P^*||Q_{x_a}).
    \end{align}
    Since $x_a\neq x_b$ and $P\in D_{x_b}$ were arbitrarily chosen, this implies:
    \begin{align}
        &\underset{n\rightarrow\infty}{\lim\sup}-\frac{1}{n}\log\big( H_{1/\alpha}(X)-I_\alpha^S(X;Y^n) \big)\notag\\
        &\leq \min_{x\neq x^\prime}\inf_{P\in\bar{D}_x}D(P||Q_{x^\prime})=\min_{x\neq x^\prime}\mathscr{C}(Q_x||Q_{x^\prime}) \label{eq:upperSibson2},
    \end{align}
    where the last step used Lemma~\ref{ChernoffLemma_journal} in Appendix~\ref{app:ancillary}.

\section{Proof for Maximal Leakage}
%
We turn to proving~(\ref{eq:Sibsonresult}) for the case $\alpha = \infty$.
While the lower bound on $I_\infty^S(X;Y^n)$ can be proven directly,
we will instead note that it
can be obtained from Lemma \ref{lowerboth} by letting $\alpha 
\rightarrow \infty$ and then $n \rightarrow \infty$.
For the upper bound, recalling the $x$-domains defined in (\ref{xDomainDef}) and (\ref{xDomainDef2}), fix $x_a \ne x_b\in\mathcal{X}$ and a $P\in D_{x_b}$ and let $\{P_n\}_{n=1}^\infty$ be a sequence such that $P_n\in\mathcal{P}_n$ for each $n$ and $P_n\rightarrow P$. Then $P_n\in D_{x_b}$ eventually and
	\begin{align}
		&I_\infty^S(X;Y^n)\leq \log\sum_{x\in\mathcal{X}}\sum_{P\in \bar{D}_x\cap\mathcal{P}_n} Q(T(P)|x)\\
		&=\log\big[|\mathcal{X}| - \sum_{x\in\mathcal{X}}\sum_{P\in \mathcal{P}_n\backslash \bar{D}_x} Q(T(P)|x)\big]\\
		&\leq\log\big[|\mathcal{X}| - \sum_{P\in \mathcal{P}_n\backslash \bar{D}_{x_a}} Q(T(P)|x_a)\big]\\
		&\leq\log\big[|\mathcal{X}| - Q(T(P_n)|x_a)\big],
	\end{align}
	eventually. Thus for sufficiently large $n$,
	\begin{align}
		&I_\infty^S(X;Y^n)\notag\\
		&\leq \log\big[|\mathcal{X}| -\frac{1}{(n+1)^{|\mathcal{Y}|}} 2^{-nD(P_n||Q_{x_a})}\big]\\
        &\leq \log\big[|\mathcal{X}|\big] -\frac{1}{(\ln 2)|\mathcal{X}|(n+1)^{|\mathcal{Y}|}} 2^{-nD(P_n||Q_{x_a})}\label{maxL_upper}
	\end{align}
    and
	\begin{align}
		&\underset{n\rightarrow\infty}{\lim\sup}-\frac{1}{n}\log\big( |\mathcal{X}| - I_\infty^S(X;Y^n) \big) \notag\\
		&\leq \underset{n\rightarrow\infty}{\lim}D(P_n||Q_{x_a})=D(P||Q_{x_a}).
	\end{align}
    Since $x_a \ne x_b$ and $P$ were arbitrary, the result follows
    by Lemma \ref{ChernoffLemma_journal} in Appendix~\ref{app:ancillary}.

\section{Proof for Arimoto}\label{sec::arimotoproof}

Note that (\ref{eq:Arimotoresult}) for the case $\alpha=1$
has already been proven. We prove the lower 
and upper bounds for the $\alpha>1$ case as follows.

\subsection{Proof of Lower Bound}

\begin{proof}
Let $|\mathcal{X}|=M$ and
\begin{align}
    \epsilon_{X|Y^n} &=\min_{f:\mathcal{Y}^n\rightarrow X}{P(X\neq f(Y^n))}\\
    &= 1-E_{Y^n}[\max_x{Q(x|Y^n)}]\\
    &\leq 1-p_{max} \text{ where } p_{max}=\max_X Q(X)\\
    &\leq 1-\frac{1}{M}.
\end{align}

For $1 < \alpha < \infty$, 
\begin{equation}
    H_\alpha(X|Y^n) \leq \log{M} - d_\alpha(\epsilon_{X|Y^n}||1-\frac{1}{M})
\end{equation}
where $d_\alpha(p||q)$ is the binary Renyi divergence (\hspace{1sp}\cite{sason_arimotorenyi_2018}, Thm. 3):
\begin{equation}
    d_\alpha(p||q) = \frac{1}{\alpha-1} \log{(p^\alpha q^{1-\alpha}+(1-p)^\alpha(1-q)^{1-\alpha})}.
\end{equation}
So, 
\begin{align}
    I_\alpha^A(X;Y^n) &= H_\alpha(X)-H_\alpha(X|Y^n)\\
    &\geq H_\alpha(X) - \log{M} + d_\alpha(\epsilon_{X|Y^n}||1-\frac{1}{M})\\
    &= H_\alpha(X) - \log{M}\notag\\
    &~~~~+\frac{1}{\alpha-1} \log \Big( \epsilon_{X|Y^n}^\alpha (1-\frac{1}{M})^{1-\alpha}\notag\\
    &~~~~+(1-\epsilon_{X|Y^n})^\alpha(\frac{1}{M})^{1-\alpha}\Big)\\
    &\geq H_\alpha(X) - \log{M} \notag\\
    &~~~~+ \frac{1}{\alpha-1} \log{((1-\epsilon_{X|Y^n})^\alpha(\frac{1}{M})^{1-\alpha})}\\
    &= H_\alpha(X) +\frac{\alpha}{\alpha-1}\log{(1-\epsilon_{X|Y^n})}
\end{align}
Hence,
\begin{align}
    I_\alpha^A(X;Y^n) - H_\alpha(X) &\geq \frac{\alpha}{\alpha-1}\log{(1-\epsilon_{X|Y^n})}\\
    \intertext{which gives}
    \frac{\alpha-1}{\alpha}[H_\alpha(X)-I_\alpha^A(X;Y^n)] &\leq \log{\frac{1}{1-\epsilon_{X|Y^n}}}. \label{arimoto1}
\end{align}
For $0 < \epsilon \le 1/2$,
\begin{align}
    \log{\frac{1}{1-\epsilon}} &= \log(1+\frac{\epsilon}{1-\epsilon})\\
    &\leq \frac{\epsilon}{1-\epsilon} \frac{1}{\ln{2}}\\
    &\leq \frac{2\epsilon}{\ln{2}}. \label{arimoto2}
\end{align}
For all sufficiently large $n$, we have $\epsilon_{X|Y^n} \le 1/2$
by the unique row assumption.
Thus, combining (\ref{arimoto1}) and (\ref{arimoto2}), for
all $1 < \alpha < \infty$,
\begin{align}
    \frac{2\epsilon_{X|Y^n}}{\ln{2}} & \geq \frac{\alpha-1}{\alpha}[H_\alpha(X)-I_\alpha^A(X;Y^n)]\\
    -\frac{1}{n}\log(\frac{2\epsilon_{X|Y^n}}{\ln{2}}) & \leq -\frac{1}{n}\log(\frac{\alpha-1}{\alpha}[H_\alpha(X)-I_\alpha^A(X;Y^n)]), \label{eq:Arimotolowerpartial}
\end{align}
and, taking $\alpha \rightarrow \infty$ in (\ref{eq:Arimotolowerpartial}),
\begin{equation}
    -\frac{1}{n}\log(\frac{2\epsilon_{X|Y^n}}{\ln{2}}) \leq -\frac{1}{n}\log(H_\infty(X)-I_\infty^A(X;Y^n)).
\end{equation}

Note that $\epsilon_{X|Y^n}$ is bounded as~\cite[Thm.~15]{sason_arimotorenyi_2018}
\begin{equation}
    \epsilon_{X|Y^n} \leq (M-1)\exp{\left(-\min_{x \neq x^\prime}\mathscr{C}(Q_x^n||Q_{x^\prime}^n)\right)}.
\end{equation}
    Then, using Lemma \ref{nChernLemma} in Appendix~\ref{app:ancillary}, for any $\alpha \in (1,\infty]$,
\begin{align}
    &\min_{x\neq x^\prime}\mathscr{C}(P_x||P_{x^\prime})  \notag\\
    &~~~~\leq \liminf_{n\rightarrow\infty}-\frac{1}{n}\log{[H_\alpha(X)-I_\alpha^A(X;Y^n)]}
\end{align}

\end{proof}

\subsection{Proof of Upper Bound}
\begin{proof}

    For $\alpha \in [0,\infty]$~\cite[(165)]{sason_arimotorenyi_2018},
\begin{equation}
    H_\alpha(X|Y^n) \geq \log\frac{1}{1-\epsilon_{X|Y^n}}.
\end{equation}
Thus,
\begin{align}
    I_\alpha^A(X;Y^n) & \leq H_\alpha(X) - \log\frac{1}{1-\epsilon_{X|Y^n}}\\
    \log\frac{1}{1-\epsilon_{X|Y^n}} & \leq H_\alpha(X) - I_\alpha^A(X;Y^n)\\
    \frac{\epsilon_{X|Y^n}}{\ln 2} & \le H_\alpha(X) - I_\alpha^A(X;Y^n)
    \intertext{and so}
    \limsup_{n\rightarrow\infty} -\frac{1}{n}\log{\epsilon_{X|Y^n}} & \geq \limsup_{n\rightarrow\infty} -\frac{1}{n}\log[H_\alpha(X) - I_\alpha^A(X;Y^n)].
\end{align}
It remains to show that 
\begin{equation}
    \limsup_{n \rightarrow \infty} - \frac{1}{n} \log
    \epsilon_{X|Y^n}  \le \min_{x \ne x^\prime} \mathscr{C}(Q_x||Q_{x^\prime}).
\end{equation}
To this end, for any $i \ne j$, we have
\begin{align}
    \epsilon_{X|Y^n} & = 
        E_{Y^n}[1-\max_x{Q(x|Y^n)}] \\
        & \ge \sum_{y^n} Q(y^n) \min(Q(x_i|y^n),Q(x_j|y^n)) \\
        & \ge \min(Q(x_i),Q(x_j)) \sum_{y^n} Q(y^n) \min\left(\frac{Q(x_i|y^n)}{Q(x_i)},\frac{Q(x_j|y^n)}{Q(x_j)}\right) \\
        & = 2 \min(Q(x_i),Q(x_j)) \epsilon_{n,i,j},
\end{align}
where
\begin{equation}
    \epsilon_{n,i,j} = \frac{1}{2} \sum_{y^n} \min(Q(y^n|x_i),Q(y^n|x_j))
\end{equation}
is the error probability for the alternative problem 
in which $X$ assumes only two values, $x_i$ and $x_j$, 
which are equally likely, and we seek to guess $X$
from $Y^n$. By \cite[Thm.~11.9.1]{cover_elements_2006},
we have
\begin{equation}
    \lim_{n \rightarrow \infty} - \frac{1}{n} \log
    \epsilon_{n,i,j} = \mathscr{C}(Q_{x_i}||Q_{x_j}).
\end{equation}
But $i$ and $j$ were arbitrary.

%
%
\end{proof}

\section{Proof for $\alpha$-Maximal Leakage}\label{sec::maxalphleak}


Note that for $\alpha=1$, $\alpha$-maximal leakage is given by regular mutual information, so that case is already proven.

\subsection{Proof of Lower Bound}
\begin{proof}
We obtain the lower bound by choosing $X\sim Q_u$, where $Q_u(X)$ denotes the uniform distribution over $\mathcal{X}$. Then
\begin{equation}
     \mathcal{L}_\alpha^{max}(X\rightarrow Y)
      = \max_{Q(X)} I_\alpha^S(X;Y^n) \geq I_\alpha^S(X;Y^n)|_{Q_u(X)}.
\end{equation}
    Then by (\ref{eq:Sibsonresult}),
\begin{equation}
    \liminf_{n\rightarrow\infty} -\frac{1}{n}\log (\log{|\mathcal{X}|} -  \mathcal{L}_\alpha^{max}(X\rightarrow Y)) \geq \min_{x\neq x^\prime}\mathscr{C}(Q_x||Q_{x^\prime})
\end{equation}
\end{proof}

\subsection{Proof of Upper Bound}
\begin{proof}
    As with the proof for Shannon capacity,
the idea is to show that the maximizing $Q(X)$ must
eventually be contained in a neighborhood of the 
uniform distribution. Over this neighborhood,
we can use Lemma~\ref{upperboth} to uniformly bound the
difference
\begin{equation}
    \log |\mathcal{X}| - \max_{Q(X)} I_{\alpha}^S(X:Y^n).
\end{equation}
First, for each $n$, let
\begin{equation}
    Q_n(X) \in \arg\max_{Q(X)} I_\alpha^S(X;Y^n).
\end{equation}
We have~\cite[Ex.~2 and Thm.~3]{Verdu:Sibson}
    \begin{equation}
    H_{1/\alpha}(X)|_{Q_n(X)} \geq I_\alpha^S(X;Y^n)|_{Q_n(X)},
    \end{equation}
    and thus, by Lemma~\ref{lowerboth},
\begin{align}
    H_{1/\alpha}(X)|_{Q_n(X)} 
    &\geq I_\alpha^S(X;Y^n)|_{Q_u(X)}\\
    &\geq H_{1/\alpha}(X)|_{Q_u(X)}  \notag\\
    &~~~~-\frac{\alpha}{(\alpha-1)\ln 2}(\Gamma_{n}+\frac{\Gamma_{n}^2}{2(1-\Gamma_{n})}).
\end{align}
    Then, 
\begin{align}
    &H_{1/\alpha}(X)|_{Q_n(X)} \geq H_{1/\alpha}(X)|_{Q_u(X)}  \notag\\
    &~~~~ -\frac{\alpha}{(\alpha-1)\ln 2}(\Gamma_n+\frac{\Gamma_n^2}{2(1-\Gamma_n)})\\
    &H_{1/\alpha}(X)|_{Q_u(X)} - H_{1/\alpha}(X)|_{Q_n(X)} \notag\\
    &~~~~ \leq \frac{\alpha}{(\alpha-1)\ln 2}(\Gamma_n+\frac{\Gamma_n^2}{2(1-\Gamma_n)})\\
    &D_{1/\alpha}(Q_n(X)||Q_u(X)) \notag\\
    &~~~~\leq \frac{\alpha}{(\alpha-1)\ln 2}(\Gamma_n+\frac{\Gamma_n^2}{2(1-\Gamma_n)})\equiv \epsilon_n,
\end{align}
where we have used the fact that $H_{1/\alpha}(X)|_{Q_u(X)} - H_{1/\alpha}(X)|_{Q_n(X)} = D_{1/\alpha}(Q_n(X)||Q_u(X))$.
Note that $\lim_{n \rightarrow \infty} \epsilon_n = 0$.
    Then, using the R\'{e}nyi version of Pinsker's Inequality (\hspace{1sp}\cite[Thm.~31]{van_erven_renyi_2014}),
\begin{align}
    D_{1/\alpha}(Q_u(X)||Q_n(X)) &\geq \frac{2}{\alpha}\sup_{A} |Q_n(A) - Q_u(A)|^2 \\
         &\geq \frac{2}{\alpha}\sup_x |Q_n(x) - Q_u(x)|^2 \\ 
    \intertext{and so}
    \epsilon_n  &\geq \frac{2}{\alpha}\sup_x |Q_n(x) - Q_u(x)|^2.
\end{align}
It also follows that, under this constraint,
\begin{align}
    \epsilon_n & \geq \frac{2}{\alpha}(Q_u(x) - \min_{x^\prime}{Q_n(x^\prime)})^2\\
    \sqrt{\frac{\alpha\epsilon_n}{2}} & \geq Q_u(x) - \min_{x^\prime}{Q_n(x^\prime)}\\
    \min_{x^\prime}{Q_n(x^\prime)} & \equiv Q_{\min,n}(X) \geq \frac{1}{|\mathcal{X}|} - \sqrt{\frac{\alpha\epsilon_n}{2}} \label{malEps1}
\end{align}
and similarly,
\begin{equation}
    \max_{x^\prime}{Q_n(x^\prime)} \equiv Q_{\max,n}(X) \leq \frac{1}{|\mathcal{X}|} + \sqrt{\frac{\alpha\epsilon_n}{2}} \label{malEps2}
\end{equation}
    Let $A_n$ be the set of distributions over $X$ that satisfy both~(\ref{malEps1}) 
    and (\ref{malEps2}) and note that $Q_n \in A_n$ eventually. Recalling~(\ref{eq:Fdef}), define
\begin{equation}\label{eq:FPQ}
        F(x,P,\tilde{Q}) = \tilde{Q}(x)Q(T(P)|x)^\alpha,
\end{equation}
    where we now indicate the dependence on the input distribution $\tilde{Q}(x)$. Similarly, we let $\{E_{x_i,\tilde{Q}}^{(n)}\}$ 
    be a partition of $\mathcal{P}_n$ such that $P \in E_{x,\tilde{Q}}^{(n)}$ implies
    $F(x,P, \tilde{Q}) =  \max_{x'} F(x',P,\tilde{Q})$ and we let $x_1(P,\tilde{Q})$, $x_2(P,\tilde{Q})$, \ldots,
    denote the letters of $\mathcal{X}$ in decreasing order of (\ref{eq:FPQ}).
    By Lemma~\ref{upperboth}, we have, eventually
\begin{align}
    &\max_{\tilde{Q}}I_\alpha^S(X;Y^n)\notag\\
    &= \max_{\tilde{Q}\in A_n}I_\alpha^S(X;Y^n)\notag\\
       &\leq \max_{\tilde{Q}\in A_n} H_{1/\alpha}(X)+\frac{\alpha}{(\alpha-1)\ln 2}\frac{1}{\underset{x\in\mathcal{X}}{\sum}\tilde{Q}(x)^{1/\alpha}}\sum_{x\in\mathcal{X}}\sum_{P\not\in E_{x,\tilde{Q}}^{(n)}}\notag\\
       &\phantom{====}\cdot(F(x_1(P,\tilde{Q}),P,\tilde{Q})^{1/\alpha-1}-F(x,P,\tilde{Q})^{1/\alpha-1})F(x,P,\tilde{Q})).\label{eq:maxAlphLower}
\end{align}
Fix $x_a \ne x_b$ and $P^* \in D_{x_b}$ and let $P_n$ be a sequence of types 
converging to $P^*$. Then for all sufficiently large $n$, we have that
$P_n \in E_{x_b,\tilde{Q}}^{(n)}$ for all $\tilde{Q} \in A_n$. Then because the summands in (\ref{eq:maxAlphLower}) are nonpositive, we have
\begin{align}
    &\max_{\tilde{Q}\in A_n}I_\alpha^S(X;Y^n)\notag\\
    &\leq \max_{\tilde{Q}\in A_n} H_{1/\alpha}(X)+\frac{\alpha}{(\alpha-1)\ln 2}\frac{1}{\underset{x\in\mathcal{X}}{\sum}\tilde{Q}(x)^{1/\alpha}}\notag\\
    &\phantom{=}\cdot(F(x_1(P_n,\tilde{Q}),P_n,\tilde{Q})^{1/\alpha-1}-F(x_a,P_n, \tilde{Q})^{1/\alpha-1})F(x_a,P_n, \tilde{Q})).
\end{align}
Note that, eventually, $x_1(P_n,\tilde{Q}) = x_b$ for all $\tilde{Q} \in A_n$ and
$F(x_b,P_n,\tilde{Q})^{1/\alpha-1} < \frac{1}{2} F(x_a, P_n, \tilde{Q})^{1/\alpha - 1}$
for all $\tilde{Q} \in A_n$. The remainder of the argument proceeds 
analogously to the Sibson proof. Eventually, we have
\begin{align}
    &\max_{\tilde{Q}\in A_n}I_\alpha^S(X;Y^n)\notag\\
    & \le \max_{\tilde{Q}\in A_n} H_{1/\alpha}(X) - \frac{1}{2} \frac{\alpha}
       {(\alpha - 1) \ln 2} \cdot \frac{1}{\sum_{x \in \mathcal{X}}
        \tilde{Q}(x)^{1/\alpha}} \\
    &      \phantom{\le \max} \cdot F(x_a,P_n,\tilde{Q})^{1/\alpha} \\
    & \le \max_{\tilde{Q}\in A_n} H_{1/\alpha}(X) - \frac{1}{2} \frac{\alpha}
     {(\alpha - 1) \ln 2} \cdot \frac{1}{|\mathcal{X}| \left(\frac{1}{|\mathcal{X}|} +        \sqrt{\frac{\alpha \epsilon_n}{2}}\right)^{1/\alpha}} \\
    & \phantom{\le \max} \cdot \left(\frac{1}{|\mathcal{X}|} - \sqrt{\frac{\alpha
        \epsilon_n}{2}}\right)^{1/\alpha}
    \frac{1}{(n+1)^{|\mathcal{Y}|}}
             2^{-n D(P_n||Q_{x_a})} \\
    & \le \log |\mathcal{X}| - \frac{1}{2} \frac{\alpha}
    {(\alpha - 1) \ln 2}  \frac{1}{|\mathcal{X}| \left(\frac{1}{|\mathcal{X}|} +
      \sqrt{\frac{\alpha \epsilon_n}{2}}\right)^{1/\alpha}} \\
    &    \phantom{\le \log |\mathcal{X}| - \mbox{ }} \cdot
          \left(\frac{1}{|\mathcal{X}|} - \sqrt{\frac{\alpha
        \epsilon_n}{2}}\right)^{1/\alpha}
        \frac{1}{(n+1)^{|\mathcal{Y}|}} 2^{-n D(P_n||Q_{x_a})}.
\end{align}
This implies that
\begin{align}
    \lim_{n \rightarrow \infty} - \frac{1}{n}
       \log \left( \log |\mathcal{X}| - 
          \max_{\tilde{Q}(X)} I_\alpha^S(X;Y^n) \right)
       \le \min_{x \ne x'} \mathscr{C}(Q_x||Q_{x'})
\end{align}
by Lemma~\ref{ChernoffLemma_journal} in Appendix~\ref{app:ancillary},
which implies the result for 
$1 < \alpha < \infty$. The $\alpha = \infty$ case
follows from (\ref{eq:Sibsonresult}) since $I_\infty^S(X;Y^n)$
does not depend on $Q(X)$, and 
$H_{1/\alpha}(X) = \log |\mathcal{X}|$ in that case.
\end{proof}

\section*{Acknowledgment}

This research was supported by the US National Science Foundation under grants
1704443, 1815893, and 1934985, and by the US Army Research
Office under grant W911NF-18-1-0426.

\appendix
\section{An Ancillary Lemma}
\label{app:ancillary}

Recall that $Q_x$ denotes the distribution of $Y$ given $x$, and for 
any $P\in\mathcal{P}$, $x_k(P)$ denotes $x\in\mathcal{X}$ such that $D(P||Q_x)$ is the $k^{th}$ smallest relative 
entropy across all elements of $\mathcal{X}$. 

\begin{lemma}\label{ChernoffLemma_journal}
    \begin{equation}
        \inf_{P\in\mathcal{P}} D(P||Q_{x_2(P)}) = \min_{x\neq x^\prime} \mathscr{C}(Q_x||Q_{x^\prime}),
    \end{equation}
    where both quantities may be infinite.
\end{lemma}
\begin{proof}
\label{sec:lemmaproof}
    We will separately prove that
    \begin{align}
        \label{eq:app:upper}
        \inf_{P\in\mathcal{P}} D(P||Q_{x_2(P)}) & \leq \min_{x\neq x^\prime} \mathscr{C}(Q_x||Q_{x^\prime}) \\
        \intertext{and}
        \inf_{P\in\mathcal{P}} D(P||Q_{x_2(P)}) & \geq \min_{x\neq x^\prime} \mathscr{C}(Q_x||Q_{x^\prime}).
\end{align}
To prove the upper bound, fix $x\neq x^\prime$ and consider $P_\lambda(y) = P_\lambda(Q_x,Q_{x^\prime},y)$
    as defined in (\ref{eq:tilteddiv}).
    Choose $\lambda^*$ such that $D(P_{\lambda^*}||Q_x)=D(P_{\lambda^*}||Q_{x^\prime})$. Then, certainly
\begin{equation}
    D(P_{\lambda^*}||Q_{x_2(P_{\lambda^*})})\leq \mathscr{C}(Q_x||Q_{x^\prime})
\end{equation}
since we know of two $X$-values whose corresponding $Q(Y|X)$ distributions are equidistant to $P_{\lambda^*}$,
from which~(\ref{eq:app:upper}) follows.

For the lower bound, we first define subsets of $\mathcal{P}$:
\begin{equation}
    E_x=\{P\in\mathcal{P}\ |\ D(P||Q_x)\leq \mathscr{C}(Q_x||Q_{x^\prime})\}
\end{equation}
\begin{equation}
    E_{x^\prime}=\{P\in\mathcal{P}\ |\ D(P||Q_{x^\prime})\leq \mathscr{C}(Q_x||Q_{x^\prime})\}
\end{equation}
    Note that $E_x$ and $E_{x^\prime}$ are convex sets since $D(\cdot||\cdot)$ is convex and that $P_{\lambda^*}$ achieves the minimum distance to $Q_{x^\prime}$ in $E_x$ and the minimum distance to $Q_x$ in $E_{x^\prime}$~\cite[Sec.~11.9]{cover_elements_2006}.

Choose any $P\in\mathcal{P}$. There are three cases to consider, depending on the location of $P$ in $\mathcal{P}$-space.

\noindent\textbf{Case 1:} $P\notin E_x$ and $P\notin E_{x^\prime}$. By construction, $D(P||Q_x)\geq \mathscr{C}(Q_x||Q_{x^\prime})$ and $D(P||Q_{x^\prime})\geq \mathscr{C}(Q_x||Q_{x^\prime})$.

    \noindent\textbf{Case 2:} $P\in E_x$. Using the Pythagorean theorem for relative entropy~\cite[Thm.~11.6.1]{cover_elements_2006},
\begin{equation}
    D(P||Q_{x^\prime})\geq D(P||P_{\lambda^*}) + D(P_{\lambda^*}||Q_{x^\prime})
\end{equation}

\noindent\textbf{Case 3:} $P\in E_{x^\prime}$. By the same argument,
\begin{equation}
    D(P||Q_x)\geq D(P||P_{\lambda^*}) + D(P_{\lambda^*}||Q_x)
\end{equation}

Hence, for any $P\in\mathcal{P}$,
\begin{equation}
    \max\{D(P||Q_x),D(P||Q_{x^\prime})\} \geq \mathscr{C}(Q_x||Q_{x^\prime})
\end{equation}

\noindent Since $D(P||Q_{x_2(P)}) = \underset{x\neq x^\prime}{\min}\ \max\{D(P||Q_x),D(P||Q_{x^\prime})\}$,

\begin{equation}
    \inf_{P\in\mathcal{P}} D(P||Q_{x_2(P)}) \geq \min_{x\neq x^\prime}\mathscr{C}(Q_x||Q_{x^\prime}).
\end{equation}
\end{proof}

The following result is standard; we provide a proof
for completeness.

\begin{lemma} \label{nChernLemma}
For any discrete distributions $P_1$ and $P_2$ on a common alphabet $\mathcal{X}$,
\begin{equation}
    \mathscr{C}(P_1^n||P_2^n) = n\mathscr{C}(P_1||P_2)
\end{equation}
\end{lemma}

\begin{proof}
    From~(\ref{eq:Chernoff:alt}),
\begin{equation}
    \mathscr{C}(P_1||P_2) = -\min_{0\leq\lambda<1}{\log{\left(\sum_x P_1(x)^\lambda P_2(x)^{1-\lambda}\right)}}.
\end{equation}
Furthermore,
\begin{align}
    &\log{\left(\sum_{x^n}P_1(x^n)^\lambda P_2(x^n)^{1-\lambda}\right)}\\
    &= \log\left(\sum_{x_1}\sum_{x_2}...\sum_{x_n}\prod_i^{n} P_1(x_i)^\lambda P_2(x_i)^{1-\lambda}\right)\\
    &= \log\left(\prod_i^{n}\sum_{x_i} P_1(x_i)^\lambda P_2(x_i)^{1-\lambda}\right)\\
    &= \log{\left(\sum_{x\in\mathcal{X}}P_1(x)^\lambda P_2(x)^{1-\lambda}\right)^n}.
\end{align}
Hence,
\begin{align}
    \mathscr{C}(P_1^n||P_2^n) &= -\min_{0\leq\lambda<1}{\log{\left(\sum_{x^n} P_1(x^n)^\lambda P_2(x^n)^{1-\lambda}\right)}}\\
    &= -\min_{0\leq\lambda<1} n\log{\left(\sum_{x\in\mathcal{X}}P_1(x)^\lambda P_2(x)^{1-\lambda}\right)}\\
    &= n\mathscr{C}(P_1||P_2). 
\end{align}
\end{proof}

\section{Data Processing for Arimoto Mutual Information}
\label{app:arimotocounter}

As a generalization of Shannon conditional entropy, Arimoto-R\'{e}nyi 
conditional entropy satisfies a number of desirable 
properties.
In particular, the rule that conditioning cannot increase entropy
carries over to the Arimoto-R\'{e}nyi version \cite{arimoto_MI_1975},
\cite[Thm.~2]{Fehr:Arimoto}, \cite[Corr.~1]{Aishwarya:Arimoto},
\cite[Prop.~2]{Arikan:Guessing}:
\begin{equation}
    H_\alpha(X|Y,Z) \le H_\alpha(X|Y).
\end{equation}
It follows from the definition of Arimoto mutual
information that a ``right-hand'' data processing inequality
therefore holds: if $X \leftrightarrow Y \leftrightarrow Z$
form a Markov chain, then
\begin{equation}
    I_\alpha^A(X;Z) \le I_\alpha^A(X;Y).
\end{equation}
To reduce our problem to an instance satisfying the distinct row 
assumption using the technique in Section~\ref{sec:result}, we require 
a ``left-hand'' version of the inequality, i.e.,
\begin{equation}
    I_\alpha^A(X;Z) \le I_\alpha^A(Y;Z)?
\end{equation}
In fact, this inequality can fail dramatically.

\begin{prop}
    For any $1 < \alpha < \infty$, there exist random variables
    $X$, $Y$, and $Z$ such that $X \leftrightarrow Y \leftrightarrow Z$
    and $Y \leftrightarrow X \leftrightarrow Z$ with $I_\alpha^A(X;Z)$
    being arbitrarily small and $I_\alpha^A(Y;Z)$ being arbitrarily large.
\end{prop}

\begin{proof}
Fix positive integers $K$ and $L$ and $0 < \epsilon < 1/L$. Let $Y$
    and $Z$ be jointly distributed as
    \begin{align}
        P(Y = i) & = \begin{cases}
            \epsilon & \text{if $i \in \{1,\ldots,L\}$} \\
            \frac{1 - L\epsilon}{K} & \text{if $i \in \{L + 1,\ldots,L + K\}$} \\
        \end{cases} \\
        P(Z = j|Y = i) & = 
        \begin{cases}
            1 & \text{if $j = i$ and $i \in \{1,\ldots,L\}$} \\
            \frac{1}{L} & \text{if $i \in \{L+1,\ldots,L+K\}$}  \\
            0 & \text{otherwise}.
        \end{cases}
    \end{align}
    We then couple $X$ to $Y$ and $Z$ via 
    \begin{equation}
        X = \min(Y,L+1).
    \end{equation}
    From~(\ref{eq:def:arimoto}), as $\epsilon \rightarrow 0$, we have that
    $I_\alpha^A(X;Z) \rightarrow 0$. Fix $\epsilon$ so that
     $I_\alpha^A(X;Z)$ is as small as desired. If we then
     let $K \rightarrow \infty$, we have
    \begin{equation}
        I_\alpha^A(Y;Z) \rightarrow \frac{\alpha}{\alpha - 1} \log L.
    \end{equation}
    But $L$ was arbitrary.
\end{proof}

For Sibson mutual information and $\alpha$-maximal leakage,
we could reduce our problem to one satisfying the unique
row assumption by dividing $\mathcal{X}$ into equivalence classes 
based on $P_{Y|X}(\cdot|x)$ and assigning to a ``leader'' 
realization in each equivalence class the probability of all of the 
$x$ realizations in that class. This approach fails for Arimoto
mutual information, due to the above result,
but the reduction is still possible
if one accounts for the exponential tilting of $P(x)$ in 
(\ref{eq:def:arimoto}).

\begin{prop}
    Fix $\alpha > 0$. If $(X,Y)$ does not satisfy the unique
    row assumption then there exists $\tilde{X}$ such that
    \begin{enumerate}
        \item[(\emph{i})] The support of $\tilde{X}$ is strictly
            contained within the support of $X$;
        \item[(\emph{ii})] $P_{Y|X}(y|x) = P_{Y|\tilde{X}}(y|x)$ for
            all $x$ and $y$;
        \item[(\emph{iii})] $(\tilde{X},Y)$ satisfies the unique row
               assumption; and
        \item[(\emph{iv})] $I_\alpha^A(X;Y) = I_\alpha^A(\tilde{X};Y)$.
    \end{enumerate}
\end{prop}

\begin{proof}
    For $\alpha = 1$, this follows directly from the chain rule
      for mutual information. For $\alpha \ne 1$,
    without loss of generality, we may assume that there exists
    a $k < |\mathcal{X}|$ such that
    \begin{equation}
        P_{Y|X}(\cdot|x_j) \ne P_{Y|X}(\cdot|x_i)
    \end{equation}
    for all $1 \le i < j \le k$,
    and for all $k < j \le |\mathcal{X}|$ there exists $1 \le i \le k$
    such that
    \begin{equation}
        P_{Y|X}(y|x_j) = P_{Y|X}(y|x_i) \ \text{for all $y$}.
    \end{equation}
    That is, the first $k$ rows of $P_{Y|X}$, viewed as
    a stochastic matrix, are unique, and every other row
    is a copy of one of those $k$ rows. For each $1 \le i \le k$,
    define the set of $X$ realizations
    \begin{equation}
        C_i = \left\{ x \in \mathcal{X} :
        P_{Y|X}(y|x) = P_{Y|X}(y|x_i) \ \text{for all $y$}\right\},
    \end{equation}
    and note that $C_1, \ldots, C_k$ are nonempty and form
    a partition of $\mathcal{X}$. Define $\tilde{X}$ to have
    support $\{x_1,\ldots,x_k\}$ with marginal distribution
    \begin{equation}
        P(\tilde{X} = x_i) = \frac{1}{\Gamma}\left(\sum_{x \in C_i}
                  P(X = x)^\alpha\right)^{1/\alpha},
    \end{equation}
    where
    \begin{equation}
        \Gamma = \sum_{i = 1}^k \left(\sum_{x \in C_i}
                  P(X = x)^\alpha\right)^{1/\alpha}.
    \end{equation}
    Define the joint distribution between $\tilde{X}$ and
    $Y$ through (\emph{ii}). Then (\emph{i})-(\emph{iii}) clearly 
      hold and we have
    \begin{align}
        & I_\alpha^A(X;Y) \\
            & = \frac{\alpha}{\alpha - 1} \log
                 \sum_y \left(\frac{\sum_{i = 1}^k
                   \sum_{x \in C_i} P(x)^\alpha
                    P(y|x)^\alpha}{\sum_{i = 1}^k
                    \sum_{x \in C_i} P(x)^\alpha}
                       \right)^{1/\alpha} \\
            & = \frac{\alpha}{\alpha - 1} \log
                 \sum_y \left(\frac{\sum_{i = 1}^k
                   \sum_{x \in C_i} (P(x)^\alpha/\Gamma^\alpha)
                    P(y|x)^\alpha}{\sum_{i = 1}^k
                    \sum_{x \in C_i} (P(x)^\alpha/\Gamma^\alpha)}
                       \right)^{1/\alpha} \\
            & = \frac{\alpha}{\alpha - 1} \log
                 \sum_y \left(\frac{\sum_{i = 1}^k
                 P(\tilde{X} = x_i)^\alpha
                 P(y|x)^\alpha}{\sum_{i = 1}^k
                 P(\tilde{X} = x_i)^\alpha}
                   \right)^{1/\alpha} \\
                   & = I_\alpha^A(\tilde{X};Y).
    \end{align}
\end{proof}

\bibliographystyle{IEEEtran}
\bibliography{references}

\end{document}